\documentclass[runningheads]{llncs}
\usepackage[T1]{fontenc}
\usepackage{graphicx}
\begin{document}
\title{Low-Resource Mongolian Speech Synthesis Based on Automatic Prosody  Annotation}
%
%\titlerunning{Abbreviated paper title}
% If the paper title is too long for the running head, you can set
% an abbreviated paper title here
%
\author{Xin Yuan\orcidID{0000-0002-1298-4399} \and Robin Feng\orcidID{0000-0002-8949-7214}\and Mingming Ye\orcidID{0000-0002-0703-5666}}
\authorrunning{X. Yuan et al.}
\titlerunning{Low-Resource Mongolian TTS Based on Automatic Prosody  Annotation}
%
% First names are abbreviated in the running head.
% If there are more than two authors, 'et al.' is used.
%
\institute{VXI, China\\ \email{\{xin.yuan, robin.feng, mingming.ye\}@vxichina.com}}
\maketitle              % typeset the header of the contribution
\begin{abstract}
While deep learning-based text-to-speech (TTS) models such as VITS have shown excellent results, they typically require a sizable set of high-quality <text, audio> pairs to train, which is expensive to collect. So far, most languages in the world still lack the training data needed to develop TTS systems. This paper proposes two improvement methods for the two problems faced by low-resource Mongolian speech synthesis\footnote{This paper is aimed at NCMMSC2022 special topic "Mongolian Text-to-Speech Challenge under Low-Resource Scenario". Challenge is introduced in detail can refer to the following url: http://mglip.com/challenge/NCMMSC2022-MTTSC/index.html}: a) In view of the lack of high-quality <text, audio> pairs of data, it is difficult to model the mapping problem from linguistic features to acoustic features. Improvements are made using pre-trained VITS model and transfer learning methods. b) In view of the problem of less labeled information, this paper proposes to use an automatic prosodic annotation method to label the prosodic information of text and corresponding speech, thereby improving the naturalness and intelligibility of low-resource Mongolian language. Through empirical research, the N-MOS of the method proposed in this paper is 4.195, and the I-MOS is 4.228.

\end{abstract}
\section{Introduction}
With the rapid development of deep learning, the speech synthesis system based on a large number of corpus training has achieved the effect comparable to human voice\cite{tan2021survey}. Today, speech synthesis technology is widely used in people's production and life scenarios, such as intelligent voice outbound calls, audiobooks, and intelligent voice assistants. However, speech synthesis still faces many difficulties in low-resource scenarios, and cannot achieve particularly good results. Specifically, low-resource scenarios can be divided into the following two aspects for discussion.

One aspect is that the corpus of speech synthesis is small, that is, the paired text and recording data are relatively small, and the specific performance is that the total duration of the recording is shorter. Typically, training a high-quality speech synthesis system requires about 10+ hours of recorded data. However, the Mongolian data in this competition is only 2 hours. The lack of corpus mainly affects the sound quality of the synthesized speech. In response to this problem, scholars mainly improve it through cross-language transfer learning. Pre-training TTS models for resource-rich languages can help resource-poor languages with discourse mapping\cite{tu2019end,de2020efficient}.

Another aspect is that there is less annotation information in the speech synthesis corpus. Building a high-prosody speech synthesis system usually requires building more prosodic information in the speech synthesis front-end. For Mandarin scenarios, prosodic information such as prosodic word (PW) prosodic phrase (PPH) and intonational phrase (IPH) is used to improve the overall prosody of the synthesized speech. Mongolian can also construct corresponding prosodic identification information according to its linguistic structure\cite{liu2020exploiting}. However, the Mongolian data in this competition has no relevant annotation information about prosody, and there is no alignment information between text and recording, which greatly increases the difficulty of building a high-prosody speech synthesis system. In response to this problem, scholars mainly improve it by pre-training text models. Although paired text and speech data are difficult to obtain, pure text data can be easily obtained, and language understanding or speech generation capabilities can be improved through self-supervised pre-training methods\cite{chung2019semi,wang2015word}.

For this competition, we propose to use a pre-trained speech synthesis backend and a pre-trained prosodic labeling system to solve the problem of low-resource scenarios. Specifically, for the problem of small corpus. We first trained the VITS\cite{kim2021conditional} speech synthesis backend model using the multi-speaker dataset VCTK\cite{veaux2016superseded}. Then, on the basis of the pre-trained model, the Mongolian data is used for transfer learning to obtain the final model. Aiming at the problem of less labeled information, this paper proposes to use the Automatic Prosody Annotation\cite{dai2022automatic} system to construct Mongolian prosody identification. This method simultaneously inputs speech and corresponding text to obtain the final textual prosody information. Since the same text can correspond to different prosodic markers, there will be a problem of one to many mapping. At the same time, the automatic prosodic labeling of speech and corresponding text effectively solves this problem.

In Section 2, we will introduce the method and specific training process of this paper in detail. Section 3 will introduce the data set and empirical results of this competition in detail. We then conclude our results in Section 4.

\section{Methods}
This section describes how to build a low-resource Mongolian speech synthesis system. First, related work is introduced, including the VITS end-to-end speech synthesis back-end system and automatic prosodic labeling system. Then, the construction and training process of the low-resource Mongolian speech synthesis system is carried out in detail.

\subsection{Related Works}

\subsubsection{VITS}

VITS is an end-to-end model, and the training process of the model is more convenient. At the same time, VITS adopts the variational inference with normalizing flow and adversarial training process, which improves the expressive ability of generative modeling.

For low-resource scenarios, we first train the VITS model using the multi-speaker English dataset VCTK. Transfer learning is then performed on the Mongolian dataset so that the model can more easily model acoustic features.

\subsubsection{Automatic Prosody Annotation}
Prosody information plays an important role in improving the expressiveness and naturalness of synthesized speech. However, the prosodic labeling process is time-consuming, and there is no prosodic labeling information in the dataset provided by this challenge. We utilize a automatic prosodic tagging method \cite{dai2022automatic} for Mongolian prosodic tagging. The method takes audio and text together as input, and prosodic annotation information as output. Since \cite{dai2022automatic} provides a Chinese-based pre-trained prosodic automatic annotation model, the effect of applying it directly to Mongolian will be greatly reduced. We made the following modifications:
\begin{itemize}
	\item First, convert the provided traditional Mongolian Latin counterpart to traditional Mongolian, and then convert traditional Mongolian to Cyrillic Mongolian.
	\item Second, use google translate to translate Cyrillic Mongolian into Chinese characters at the character level.
	\item Finally, input the Chinese characters and Mongolian voices into the pre-trained prosodic automatic annotation model to obtain prosodic annotation information.
\end{itemize}

\subsection{Low-Resource Mongolian Speech Synthesis}
\begin{figure}
	\includegraphics[width=\textwidth]{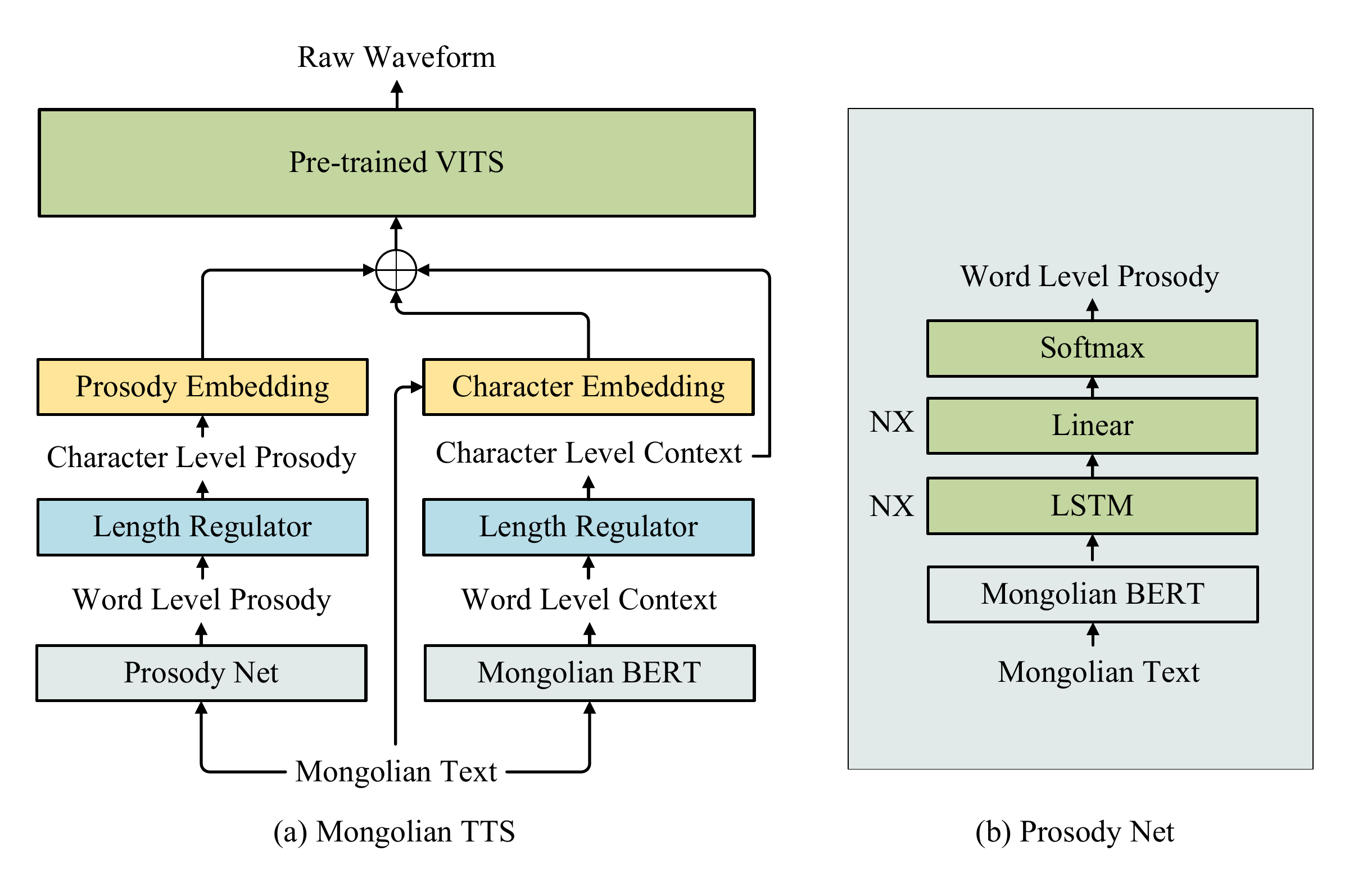}
	\caption{System diagram depicting (a) Mongolian speech synthesis rramework  and (b) Mongolian prosody prediction framework} \label{fig1}
\end{figure}

The low-resource speech synthesis system mainly includes a prosody prediction module, a pre-trained BERT model based on Cyrillic Mongolian\cite{mongolian-bert} and a VITS module, as shown in Fig.~\ref{fig1}. In this paper, the prosody prediction module is named ProsodyNet, and the pre-trained BERT based on Cyrillic Mongolian is named Mongolian BERT. ProsodyNet is mainly used to construct Word-level prosody, and Mongolian BERT is mainly used to extract Word-level context. Word-level prosodic information and contextual information are mapped to character-level through Length Regulator. Finally, all the information at the character level is input into the VITS model. 

\subsubsection{Training Process}
The ProsodyNet and VITS models are trained separately during training. We use the prosody information obtained by Automatic Prosody Annotation as ground truth to train ProsodyNet. During the transfer learning process of the VITS model using the Mongolian dataset, the input prosodic information is the real value, not the predicted value of ProsodyNet.

\subsubsection{Inference Process}
ProsodyNet is used for prosody prediction during inference.

\section{Experiments}
\subsection{Dataset}
Officially provide NCMMSC2022-MTTSC\cite{hu2022mntts} as the competition data set. The dataset was recorded by a professional female announcer whose native language is Mongolian. The dataset consists of three parts: training set, validation set and test set. Both training and validation sets consist of a metadata.csv containing text scripts and a wavs folder containing the corresponding audio. The test set consists of a metadata.csv file containing the test text script. The training set contains a total of 1000 records, the validation set contains 298 records, and the test set contains 200 records. 

\subsection{Results}
\begin{figure}
	\includegraphics[width=\textwidth]{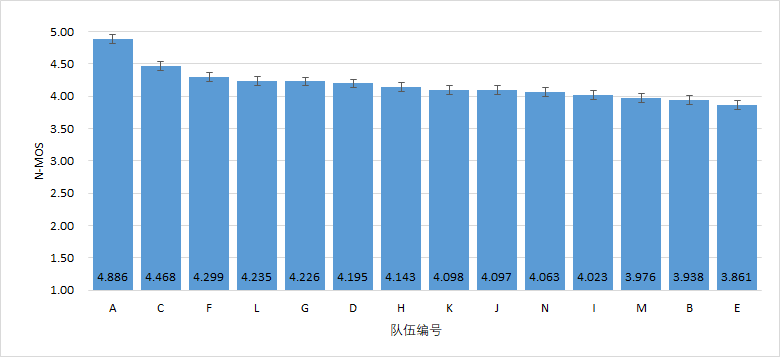}
	\caption{Naturalness Mean Opinion Score} \label{fig2}
\end{figure}

\begin{figure}
	\includegraphics[width=\textwidth]{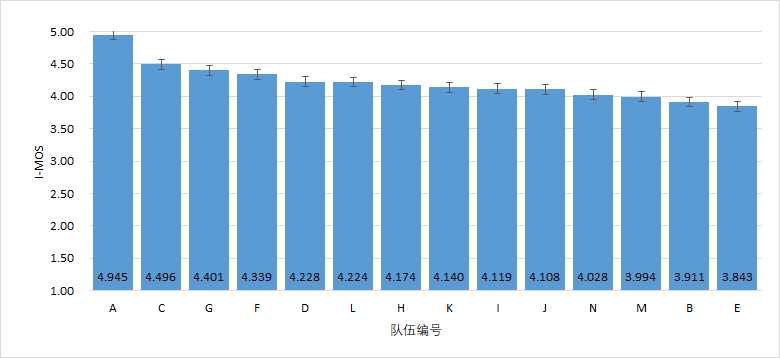}
	\caption{Intelligibility Mean Opinion Score} \label{fig3}
\end{figure}

\begin{figure}
	\includegraphics[width=\textwidth]{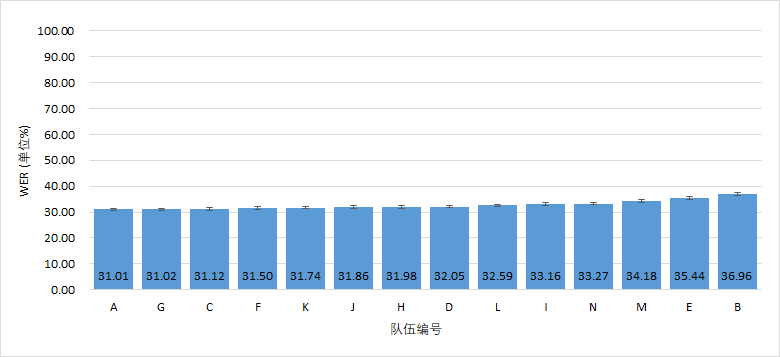}
	\caption{Speech Recognition Word Error Rate} \label{fig4}
\end{figure}

Evaluation indicators include Naturalness Mean Opinion Score (N-MOS), Intelligibility Mean Opinion Score (I-MOS) and Speech Recognition Word Error Rate (WER) . The detailed evaluation results are shown in Fig.~\ref{fig2}-Fig.~\ref{fig4}. The number of our team is D. Team A refers to the real audio evaluation results.
\footnote{Audio samples can be found in https://yuan1615.github.io/2022/10/21/MongolianTTS/}

\section{Conclusions}
This paper proposes a low-resource Mongolian speech synthesis method based on automatic prosodic labeling. The main innovations are two: a) In view of the lack of recording data, it is difficult to model the mapping of linguistic features to acoustic features. The methods of training models and transfer learning are improved. b) For the problem of less labeled information, this paper proposes to use an automatic prosodic labeling method to label the prosodic information of text and corresponding speech, so as to improve the naturalness of low-resource Mongolian language. Through empirical research, the method proposed in this paper obtains high N-MOS and I-MOS scores. However, since the pre-trained automatic prosody annotation model used in this paper is based on Chinese training, the prosody annotation results obtained here are still wrong. A possible follow-up direction for improvement is to manually annotate part of the prosody information, and then use the manually annotated prosody information to fine-tune the pre-trained automatic prosody annotation model. Using the fine-tuned model to perform prosodic annotation on all text and corresponding speech will improve the accuracy of prosodic information, thereby improving the naturalness and intelligibility of the final synthesized speech.

\newpage

%
% ---- Bibliography ----
%
% BibTeX users should specify bibliography style 'splncs04'.
% References will then be sorted and formatted in the correct style.
%
\bibliographystyle{splncs04}
\bibliography{samplepaper}
\end{document}